\documentclass[sigconf,10pt]{acmart}
\renewcommand\footnotetextcopyrightpermission[1]{}
\usepackage[english]{babel}
\usepackage[utf8]{inputenc}
\usepackage{blindtext}
\usepackage{caption}
\usepackage{subcaption}
\usepackage{listings}
\usepackage{xcolor}
\usepackage{alltt}
\usepackage{paralist}
\usepackage{enumitem}
\usepackage{comment}
\usepackage{multirow}
\usepackage{array}
\usepackage{tcolorbox}
\usepackage{stfloats}
\usepackage{tablefootnote}
\renewcommand{\ttdefault}{txtt}

\usepackage{afterpage}

\AtBeginDocument{%
  }

\usepackage{thmtools}
\usepackage[framemethod=TikZ]{mdframed}
\mdfsetup{skipabove=1em,skipbelow=0em, innertopmargin=5pt, innerbottommargin=6pt}
\theoremstyle{definition}

\definecolor{DimGray}{rgb}{0.41, 0.41, 0.41}
\definecolor{Bisque}{rgb}{242, 210, 189}
\colorlet{thmgreencolor}{DimGray}

\makeatletter

\declaretheoremstyle[
  headfont=\bfseries, bodyfont=\normalfont\itshape,
  mdframed={
    innertopmargin=2pt,
    innerleftmargin=8pt,
    innerrightmargin=8pt,
    innerbottommargin=5pt,
    linewidth=2pt,
    rightline=false, topline=false, bottomline=false,
    linecolor=black, backgroundcolor=DimGray!5,
  }
  ]{thmgraybox}
  
\makeatother

\declaretheorem[style=thmgraybox, numbered=yes, name=Takeaway]{takeaway}
\declaretheorem[style=thmgraybox, numbered=no, name=Takeaway]{takeaway2}

\AtBeginEnvironment{takeaway}{
  \captionsetup{labelfont={color=thmgraycolor}}
}

\definecolor{Gray}{gray}{0.9}
\newcolumntype{?}{!{\vrule width 1.5pt}}

\newif\ifshowcomments

\ifshowcomments
\newcommand{\mynote}[2]{\fbox{\bfseries\sffamily\footnotesize{\textbf{#1}}}
 {\small$\blacktriangleright$\textsf{\emph{#2}}$\blacktriangleleft$}}
\else
\newcommand{\mynote}[2]{}
\fi

\newcommand{\sys}{\textsc{MaLV-OS}\xspace}

\DeclareFontFamily{\encodingdefault}{\ttdefault}{\hyphenchar\font=`\-}

\begin{document}

\title{\sys: Rethinking the Operating System Architecture for Machine Learning in Virtualized Clouds
}

\author{Stella Bitchebe}
\affiliation{%
  \institution{McGill University}
  \country{Canada}
}
\email{stella.bitchebe@mcgill.ca}

\author{Oana Balmau}
\affiliation{%
  \institution{McGill University}
  \country{Canada}
}
\email{oana.balmau@mcgill.ca}

\begin{abstract}
As Machine Learning (ML) is increasingly emerging and ML models are continuously enhanced, a large body of research has employed these models to develop learned operating systems (OSes) and kernels.
The latter dynamically adapts to the job load and dynamically adjusts resources (CPU, I/O, memory, network bandwidth) allocation to respond to the actual user demand. What this work has in common is that it utilizes ML to improve kernel decisions. To this day, and to the best of our knowledge, no work has taken the opposite direction, i.e., using OS to improve ML. While some work proposes applying system-level optimizations to ML algorithms, they do not tailor the OS to adapt to the ML context. To address this limitation, we take an orthogonal approach in this paper by leveraging the OS to enhance the performance of ML models and algorithms.

We explore the path towards an ML-specialized OS, \sys. \sys rethinks the OS architecture to make it specifically tailored to ML workloads, especially in virtualized clouds, which are now widely used to run ML applications. 
\sys's envisioned architecture includes (1) a micro-kernel, \texttt{Micro-LAKE}, which allows kernel space applications to use the GPU, and (2) an MLaaS (ML as a Service) subsystem that gathers ML models to help \texttt{Micro-LAKE} with memory management and CPU scheduling. 
\sys architecture also offloads system-sensitive parts of the models to the OS, to lighten the model complexity and programming, and speed up its execution. Finally, MaLV-OS integrates an open-source GPU virtualization software, merged directly into the hypervisor. 
For more flexibility, \sys's vision is to enable the virtual machine to dynamically select MLaaS policies that can improve the performance of the model the user is running. 
Because MLaaS is designed as loadable kernel modules, the MaLV-OS architecture enables the dynamic addition of new capabilities to the MLaaS subsystem.
\end{abstract}


\maketitle

\section{Introduction and Motivation}
\label{sec:motiv}

\paragraph{\bf ML applications on non-specialized Systems}
Efficient ML models rely on two key components: (1) high-quality datasets, and (2) high-quality algorithms.
The latter has received tremendous attention~\cite{processing-sl, 10.3389/fninf.2011.00013, ml.net, doi:10.1137/S1064827595287997, 910827, 10.1021/ci100031x, 9488982} these recent years.
Besides, the entire computing stack has evolved to adapt to ML and AI workloads, with specialized hardware accelerators (like GPU, TPU, DPU, etc.), specialized software frameworks~\cite{numpy, scikit, pandas}, and specialized and enhanced algorithms~\cite{10.5281/zenodo.1082415, nuts-flow/ml, 9036908, rf-preporcessing}, that have been proposed in recent years.
However, because much less attention has been given to the system part where those algorithms run, the impact of innovations in other tiers (hardware, frameworks, etc.) cannot fully be observed, as the system is the most critical layer.
For example, recent research~\cite{speedyloader, pecan, tfdata, dali, fastflow, cachew} has shown that hardware accelerators are almost always underutilized (as low as 18$\%$ GPU usage for some workloads and frameworks, see Figure~\ref{fig:passthrought-gpu-usage}), due to the latency incurred by some operations that are performed on the OS and that slow down the end-to-end training time of the workload. They have also proven that applying even simple system-level optimizations to ML framework components, such as data loader, scheduler, or cache management, can substantially improve the training time of models.

\begin{takeaway}
    Evolved models and very fast hardware do not guarantee fast model training and inference if they run on non-specialized Systems.
\end{takeaway}

\paragraph{\bf General-purpose OSes used for ML workloads in the cloud.}
The two most popular operating systems used for ML in the cloud are Microsoft and Linux~\cite{princeton-os-cloud, wiki-os-cloud}, with a preference for Linux due to its characteristics.
Indeed, Linux is open-source, thus customizable and flexible, which makes it suitable for research.
It is lightweight and has more optimized memory management than Microsoft.
It is suitable for ML frameworks, with well-known tools like Docker and frameworks like TensorFlow, PyTorch, and Keras that are well integrated into Linux and optimized for the Linux environment.
Finally, it is offered in almost all providers’ VM instances with most distributions free of charge.
However, with datasets becoming larger and more complex, even the Linux OS is no longer adapted~\cite{7976640, 5486552, 10.1145/1807128.1807132} for ML workloads, because of the following reasons.
(1) Static scheduling decision-making, which is not adapted for streaming tasks (e.g., real-time traffic management).
(2) Memory management and memory swapping that are not adapted for large-scale and real-time data (e.g., management of irregular memory access patterns).
(3) And the lack of predictive analytics to cope with dynamic datasets.

\begin{takeaway}
    Existing general-purpose operating systems used in the cloud are not adapted for ML Workloads.
\end{takeaway}

\paragraph{\bf Emergence of ML-specialized systems}
As machine learning and artificial intelligence (AI) are continuously integrated into almost all devices, we notice the development of many systems for AI, driven by the need for specialized systems adapted to AI algorithms and infrastructures.
Among those systems, we can cite Tesla Autopilot~\cite{tesla-autopilot}, an advanced driver-assistance system that enables Tesla cars to steer, accelerate, and brake automatically.
Google Android AI~\cite{android-ai, gemini} and Apple AI~\cite{apple-ai, apple-ai-wiki}, which are intelligent systems integrated, respectively, by Google and Apple into their latest smartphones to improve and assist mobile OSes (Android and iOS, respectively) with AI tasks.
However, these are not computer OSes, and they are proprietary and domain-specific.

\begin{takeaway}
    The mobile industry has followed suit with specific systems tailored for AI, with remarkable and notable benefits. 
    We believe it is crucial to follow the trend and develop an ML-specialized cloud operating system for computers.
\end{takeaway}

\paragraph{\bf Improvement of OS components for ML training.}
Section~\ref{sec:rw} presents the state-of-the-art on (1) using OS abstractions for ML and (2) using ML techniques and algorithms to improve OS components.
With ML algorithms and models constantly evolving in number and robustness, researchers have proposed a plethora of solutions and frameworks that leverage ML to help the kernel with core tasks like memory management, process scheduling, or resource allocation. While other work has proposed applying system-level optimizations and abstractions to ML frameworks, no one, to the best of our knowledge, has proposed a whole operating system specific to ML workloads.

\begin{takeaway}
    There has been extensive work on improving single OS pieces to adapt to ML algorithms. However, to the best of our knowledge, no work has tried to aggregate all state-of-the-art improvements into a concrete OS tailored for ML applications in the cloud.
\end{takeaway}

\paragraph{\bf OS-support for GPU virtualization.}
Virtualization has become the foundation of public clouds because it provides strong isolation, resource sharing, energy cost reduction, and attractive costs to users.
Thanks to this, the Cloud is nowadays the preferred execution environment for modern applications, including ML applications.

Executing ML applications in production in virtualized clouds implies making the GPU available to virtual machines (VMs) through GPU passthrough or full GPU virtualization.
As Section~\ref{sec:overhead-full-virt} explains, the main technique used today in public clouds is PCI passthrough of the GPU.
Generally, the latter makes the GPU unavailable to the host OS and not shareable between multiple VMs. This results in low GPU utilization (Figure~\ref{fig:passthrought-gpu-usage}) and high GPU-VM costs for the client (because they need to reserve and pay for the entire physical GPU).
For example, on AWS~\cite{aws-instance-pricing}, an instance with 8$\times$A100 GPUs is priced 32.77\$-40.96\$ per hour, while an 8$\times$V100 GPUs instance costs 24.48\$.
Nevertheless, virtualization's very essence resides in its \emph{multiplexing} capability, which allows a physical resource to be shared between multiple VMs.
Therefore, like for the CPU, a full GPU virtualization technique is necessary, along with appropriate systems, to make GPU resources more flexible and affordable in public clouds.
To address this issue, NVIDIA proposes vGPU~\cite{nvidiavgpu}, a virtual GPU software that provides a subset of the physical GPU's capabilities to a VM, making the GPU sharable between multiple guest OSes.
However, NVIDIA vGPU is proprietary, limiting its broad adoption in public clouds and research.

This paper explores, in Section~\ref{sec:virt-overhead}, the overhead of current GPU virtualization techniques and shows that they are insufficient for new hardware.

\begin{takeaway}
    To enable GPU sharing and overcommitment in public virtualized clouds, we need efficient, open-source GPU full virtualization software and adequate OS support to host it.
\end{takeaway}

\section{Related Work}
\label{sec:rw}

A large body of work has taken advantage of the proliferation of ML algorithms to improve OS subcomponents such as the scheduler and the memory manager~\cite{lake,smartos,linnos,Fedorova2007OperatingSS,habibi2024aimldriven,10.1145/3409963.3410492,kleio,deepprefetch,deepmaldroid,10.1145/3373376.3378525,lynx,4085157,10.1145/3458336.3465288,10.1145/506084.506090,8664598}.
On the other hand, some work has used OS capabilities and functionalities to enhance ML frameworks~\cite{speedyloader, pecan, tfdata, dali, fastflow, cachew}.

\paragraph{\bf ML for systems.}
Kleio~\cite{kleio} is a memory page scheduler that utilizes neural network training to efficiently determine the working set (hot pages) of applications and smartly apply page replacement and swapping.
Lynx~\cite{lynx} is a prefetching mechanism developed to replace the Linux prefetching subsystem.
It learns I/O patterns and represents them in the form of a Markov chain to leverage the algorithm locality of data and prefetch them regardless of the pattern.
\cite{Fedorova2007OperatingSS} exploits reinforcement learning to propose a Linux-based scheduler that balances optimal performance, fair CPU sharing, and balanced core assignment in heterogeneous multicore processors.

\paragraph{\bf Systems for ML}
Speedyloader~\cite{speedyloader} is a recent research work that applies system optimizations to improve ML training.
Speedyloader first demonstrates that the preprocessing step of an ML training workload, which is executed on the CPU memory, can significantly impact the end-to-end training time if it is inefficiently performed. In light of this observation, Speedyloader proposes a PyTorch-based data loader that parallelizes preprocessing and training to accelerate the training and maximize GPU utilization.
Pecan~\cite{pecan} is a TensorFlow-based system for disaggregated environments that reschedules transformations applied to data in order to accelerate the model training and reduce total training costs.
DALI~\cite{dali} is the NVIDIA Data Loading Library, which proposes to use the GPU for preprocessing, to feed data to the GPU faster, in order to increase the model throughput and training time. 

\paragraph{\bf ML-based Operating Systems.}
LAKE~\cite{lake} is one of the most recent research work that brings ML to the kernel. LAKE makes the accelerator (e.g., GPU) available to kernel space applications, which allows them to make decisions based on ML algorithms more quickly.
LinnOS~\cite{linnos} is an operating system that employs deep learning to improve I/O access to SSD storage.
SmartOS~\cite{smartos} is a learning-based OS that uses reinforcement learning to dynamically allocate CPU, memory, I/O, and network bandwidth to users.
\cite{10.1145/3458336.3465288} employs ML to enable dynamic kernel self-optimizations.

While all these operating systems are driven by ML decisions, their goal is not to enhance or improve the performance of ML algorithms, but to exploit the latter to enhance the OS. \sys is orthogonal to this work and proposes an OS that is completely dedicated to ML workload execution.

\section{GPU Virtualization}
\label{sec:virt-overhead}

GPU is currently virtualized through two main techniques: full virtualization and PCI-Passthrough.
The latter gives the virtual machine (VM) direct access to the GPU and is the most used technique today in public clouds.
Few work~\cite{convgpu, nvidiavgpu, vadi, 7349172} has studied and proposed a way to fully virtualize the GPU.
However, to the best of our knowledge, none of these solutions is currently used in production. NVIDIA vGPU is the most well-known solution, yet proprietary and therefore not broadly adopted.

PCI passthrough has been extensively studied and evaluated~\cite{6820614,6969471,6427531,10.1145/3068281}.
This study has shown that GPU passthrough has almost no overhead on virtualized applications (because the VM accesses the physical GPU, not a virtual instance of it).
Since most of these studies date back over ten years,  in Section~\ref{sec:overhead-passthrough}, we assess their validity on more recent NVIDIA GPU hardware and libraries.

\subsection{GPU Passthrough}
\label{sec:overhead-passthrough}
This section presents our GPU passthrough evaluation results and exposes the limitations of passthrough for public virtualized clouds. 

\paragraph{\bf Systems and hardware.}
We have evaluated the performance of GPU passthrough on two machines.
(1) \emph{Native}, a physical server with 8$\times$V100 32GB NVIDIA GPUs, two 40-core Intel Xeon processors, 512GB of memory, and a 7TB NVMe SSD, using Ubuntu 20.04. 
(2) \emph{VM}, a KVM virtual machine configured with 256GB of memory, 64 vCPUs, and using Ubuntu 20.04.
To perform the experiments on the physical machine (\emph{Native}), we restricted it to use 256GB of memory and 64 CPUs, exactly as the VM.
Both machines run CUDA version 12.6 and use the same workload datasets and code through a shared space mounted inside the VM.

\paragraph{\bf Workloads.}
We used a diverse set of workloads from the MLPerf Training Benchmark suite~\cite{mattson2020mlperf,mlcommons-github}.
(1) \textbf{Image Segmentation}, that performs 3D segmentation of kidney tumors represented by the 29GB KiTS19 dataset~\cite{kits19,heller2019kits19}, and is trained with the 3D-UNet model~\cite{10.1007/978-3-319-46723-8_49}. 
(2) \textbf{Object Detection}, which detects objects in 2D images using the Mask R-CNN model~\cite{massa2018mrcnn} with a ResNet50 backbone, and uses a 58GB COCO dataset~\cite{cocodataset}. 
(3) \textbf{Speech Recognition}, which performs speech-to-text transcription using the RNN-T model~\cite{rnnt} trained on the 228GB LibriSpeech dataset~\cite{librispeech} containing approximately 1,000 hours of English speech. 

\paragraph{\bf Metrics.}
We measured the normalized training time and the GPU usage.
We measure the training time of each workload on both machines and compute the VM's time normalized to the Native one.
We measured the GPU usage via the \texttt{nvidia-smi} tool from the NVIDIA Management Library~\cite{nvml}.

\paragraph{\bf Training time.}

\begin{figure}[t]
    \centering
    \includegraphics[width=\columnwidth]{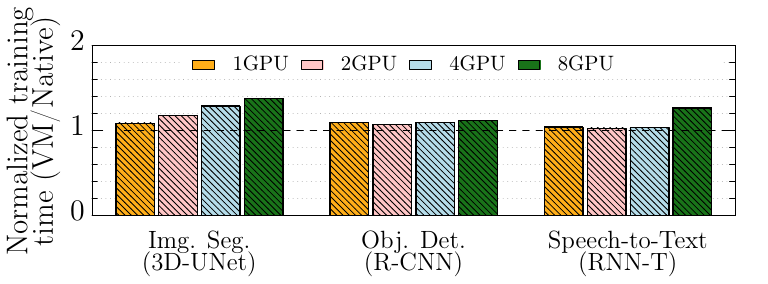}
    \vspace{-2em}
    \caption{Workload's end-to-end training time in VM (with GPU passthrough) normalized to that in Native.
    }
    \vspace{-1.3em}
    \label{fig:passthrought-training-time}
\end{figure}

Figure~\ref{fig:passthrought-training-time} depicts the training time of the virtual machine normalized to the physical server when varying the number of GPUs from 1 to 8.
First, we observe that GPU passthrough incurs only $13\%$ overhead, on average, across all workloads and GPU configurations.
However, this overhead is much higher for the image segmentation workload and reaches $37\%$ when using 8$\times$GPUs.
This is because, unlike other workloads, image segmentation has a high preprocessing load~\cite{speedyloader}.
Because preprocessing is performed on the CPU, the training time in the VM will be affected by the CPU virtualization overhead.
Second, Figure~\ref{fig:passthrought-training-time} shows that the overhead increases with the number of GPUs.
This can be explained by the fact that the more GPUs the Native machine uses, the faster the training is completed.
And because the number of GPUs does not influence the preprocessing time, preprocessing-intensive workloads would be trained much more slowly in virtual environments, even with more GPUs.

\vspace{-.5em}
\begin{takeaway2}
    Recent enhanced ML models suffer from high overhead when using passthrough on recent and faster GPU hardware. This overhead is amplified for preprocessing-intensive workloads (which make higher use of the CPU memory), supporting \sys design of offloading data loading and preprocessing to the kernel (see Section~\ref{sec:malt-os}).
\end{takeaway2}

\begin{figure*}[t]
    \centering
     \begin{subfigure}[t]{0.33\linewidth}
        \includegraphics[width=\linewidth]{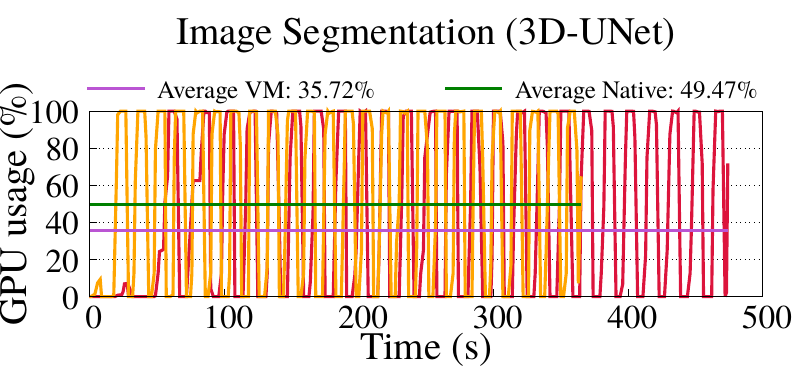}
    \vspace{-2em}
    \end{subfigure}
     \begin{subfigure}[t]{0.33\linewidth}
        \includegraphics[width=\linewidth]{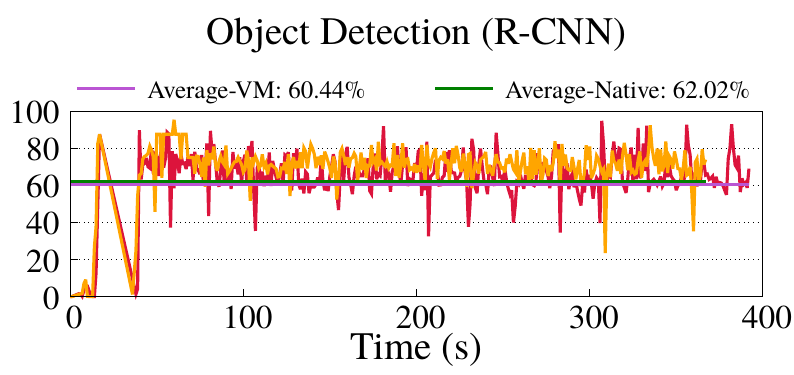}
    \vspace{-2em}
    \end{subfigure}
     \begin{subfigure}[t]{0.33\linewidth}
        \includegraphics[width=\linewidth]{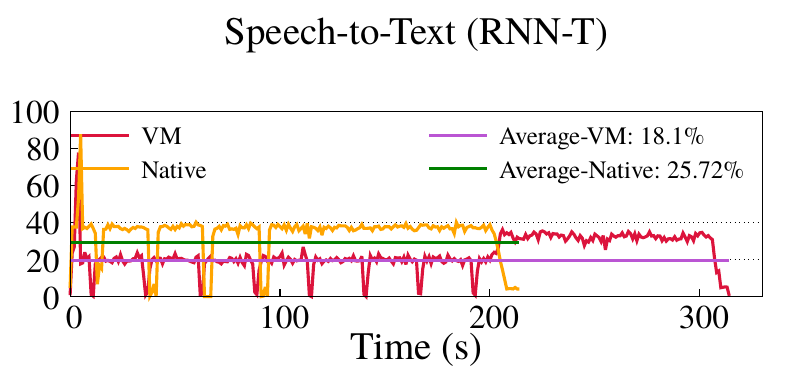}
    \vspace{-2em}
    \end{subfigure}
    \vspace{-2em}
    \caption{GPU usage in VM and Native, when using 8$\times$GPU passthrough. GPU is underutilized in both environments, with more fluctuations in the image segmentation workload, which is preprocessing-intensive.}
    \vspace{-1.em}
    \label{fig:passthrought-gpu-usage}
\end{figure*}

\paragraph{\bf GPU usage.}
GPU usage of each workload is depicted in Figure~\ref{fig:passthrought-gpu-usage} for 8$\times$GPUs.
We observe that, with both VM and Native, the GPU is underutilized, with GPU usage ranging from 18.1$\%$ for the speech workload in the VM, to 62$\%$ for object detection on Native.
However, the GPU utilization is on average 27$\%$ (2.6$\%$-42.1$\%$) lower in the virtual environment, which is non-negligible.
This observation sustains \sys vision of integrating a GPU full virtualization software (see Section~\ref{sec:malt-os}) and leveraging hardware support for virtualization to increase GPU usage through efficient GPU scheduling algorithms.

\vspace{-.5em}
\begin{takeaway2}
    For ML applications in virtual clouds to meet at least the same GPU utilization rate as in physical servers, the need for a GPU full virtualization solution is crucial.
\end{takeaway2}

~\\
\noindent While these experiments run on \emph{relatively old} GPUs (V100), we expect the difference in performance to be even higher for newer GPUs like A100 or H100.

\vspace{-.2cm}
\subsection{GPU Full Virtualization}
\label{sec:overhead-full-virt}

The main limitation of passthrough is the lack of resource sharing.
One might argue that GPU memory is too small (commonly 16GB-32GB) to provide high elasticity. However, as illustrated in Figure~\ref{fig:passthrought-gpu-usage}, the GPU is usually underutilized and can be assigned to other processes during idle periods.

In 2020, NVIDIA released the NVIDIA Virtual GPU (vGPU) software~\cite{nvidiavgpu}, the most well-known GPU full virtualization solution today.
NVIDIA vGPU is installed along with the hypervisor layer and allows for allocating multiple \emph{virtual} GPUs to a single VM and for sharing a single physical GPU between many VMs.
Because vGPU is proprietary, we have not yet evaluated it, but this is part of \sys's future work.
Before NVIDIA vGPU software, the first research work to propose a GPU virtualization solution dates back around 2009 when VMware proposed a GPU virtualization architecture~\cite{10.1145/1618525.1618534} for VMware's hosted applications.
However, besides being proprietary, VMware virtual GPU software was mainly based on a passthrough strategy.
In 2014, Y. Suzuki et al. proposed GPUvm~\cite{7349172}, an open-source GPU full- and para-virtualization support.
To the best of our knowledge, GPUvm is currently not used in production, certainly due to its significant overhead resulting from large IOMMU handling (especially for full virtualization).  
In 2017, convGPU~\cite{convgpu} went one step further by proposing a GPU management software that allows the sharing of a GPU between many containers.
Today, ML workloads are run primarily using containers such as the well-known NVIDIA Docker~\cite{nvidia-docker}.
And even inside VMs, users generally deploy containers to encapsulate their applications. 

Nowadays, for every new hardware release, a virtualization solution is immediately designed and proposed.
Indeed, there are noticeable research efforts on virtualizing other accelerators or PIM (processing-in-memory) hardware, such as DPUs (DRAM Processing Units).
vNPU~\cite{vnpu} is a virtualization system for NPUs (Neural Processing Units) that integrates a vNPU manager into the hypervisor and proposes a flexible abstraction of NPU.
The latter divides the NPU into data arrays and operation vectors.
This way, vNPU can provide the VMs with virtual vectors in the form of virtual NPU instances.
vPIM~\cite{vpim} is the first work to propose a concrete and open-source solution to virtualize DPU hardware for the cloud.
vPIM can multiplex the DPU and requires no changes from the hardware or the hypervisor.
This work emphasizes the need for an open-source GPU virtualization solution.

\section{\sys Vision}
\label{sec:malt-os}

\begin{figure}[t]
    \centering
    \includegraphics[ width=\columnwidth]{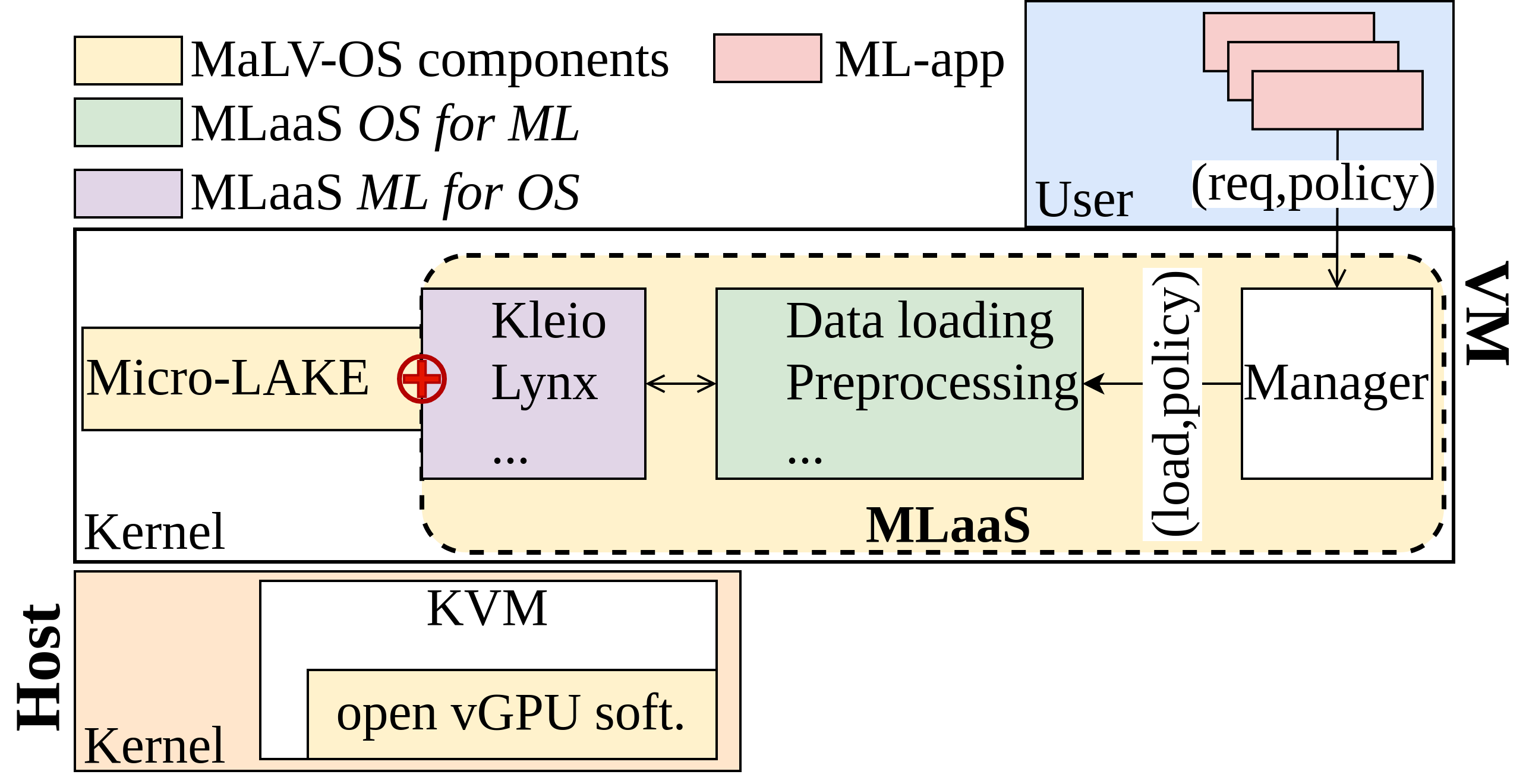}
    \caption{\sys architecture, which builds on the LAKE kernel to make the virtual GPU available to the kernel space. MLaaS stands for ML as a Service.
    The user can offload parts of the model code to \sys and request them as kernel modules (in green), loaded and unloaded by the MLaaS (ML as a Service) manager.
    }
    \label{fig:malvos}
\end{figure}

\paragraph{\bf \sys objective.}
\sys's particular focus is on integrating dynamic scheduling decisions that efficiently adapt to more dynamic and complex workloads and adequately handling dynamic datasets by using dynamic memory management.
\sys aims to achieve this goal through its envisioned architecture presented in Figure~\ref{fig:malvos}.

\paragraph{\bf \sys in a nutshell.}
Actual training, inference, and fine-tuning are performed on the GPU memory.
However, the ML framework runs on the CPU, and before feeding data to the GPU, the model generally applies preprocessing transformations that are performed on the CPU memory~\cite{9820645, speedyloader}.
Because the preprocessing step can account for more than 30\% of the end-to-end training time of an ML workload~\cite{speedyloader}, it is essential that OS memory and CPU scheduling are adequately managed to meet the ML application objectives.

\sys (Figure~\ref{fig:malvos}, yellow) rethinks the guest OS architecture to make it specific to ML applications (Figure~\ref{fig:malvos}, pink). 
It includes two main components. (1) \texttt{Micro-LAKE}, a micro-kernel (extracted from LAKE~\cite{lake}, see Section~\ref{sec:rw}) that makes the kernel space access the virtual GPU. 
\texttt{Micro-LAKE} is assisted by a kernel background model that continuously learns and provides it with updated knowledge to dynamically adjust its policies depending on the type of workload the user (Figure~\ref{fig:malvos}, light blue) is running.
(2) An ML as a Service (MLaaS) subsystem, which is composed of two main parts: \emph{OS for ML} and \emph{ML for OS}.
The latter (Figure~\ref{fig:malvos}, purple) includes models intended to assist \texttt{Micro-LAKE} with memory management and CPU scheduling, and builds upon prior work that proposes individual ML for OS improvements, such as Kleio~\cite{kleio} or Lynx~\cite{lynx} (see Section~\ref{sec:rw}).

\emph{OS for ML} (Figure~\ref{fig:malvos}, green) is designed to lighten the models (training, inference, and fine-tuning) and integrate some of their code directly into the OS as loadable kernel modules.
To achieve this, \sys architecture extends the \texttt{syscall} API to allow the user to request an MLaaS policy according to the model they are running.
The request is handled by the MLaaS manager, which loads the corresponding module.
Once the user completes its execution, the module is unloaded.
This approach relieves the programmer from managing system aspects that have a non-negligible impact on the model's performance.
As an example, with an image segmentation workload, \sys could integrate data loading and preprocessing as part of the MLaaS functionalities.
They would then be performed directly by the kernel, which would simplify the model and speed up its execution.

\paragraph{\bf \sys GPU virtualization}
\sys architecture integrates the GPU virtualization software directly as part of the hypervisor to avoid the need for an additional virtualization layer and envisions two approaches for fully virtualizing the GPU.
The first one, \emph{para-vGPU}, follows the vPIM~\cite{vpim} technique of extending the \texttt{virtio} standard and requires no hardware and no hypervisor changes. 
In \emph{para-vGPU}, \sys exposes the virtual GPU to the VM as vGPU devices by installing \texttt{virtio} drivers in the VM.
The second approach, \emph{full-vGPU}, mimics the CPU hardware-assisted virtualization~\cite{Bugnion:2013:LSS:2490483.2490506} technique and necessitates changes to the hardware and the hypervisor.
With \emph{full-vGPU}, \sys integrates the notion of \texttt{G-VMCS} (GPU-VMCS), identical to the VMCS (virtual machine control structure) for the CPU.
The hypervisor vGPU software creates and manages a distinct \texttt{G-VMCS} for each virtual GPU instance in the VM.
New control fields would be added to the hardware to guide transitions between virtual and physical modes, enabling efficient GPU sharing between host and VMs.

\paragraph{\bf \sys integration in-production.}
The design of \sys foresees an option for smooth integration in cloud environments, by requiring no changes to the hardware (with \emph{para-vGPU}), to the host (cloud provider) operating system, or to the user applications.
Indeed, \sys MLaaS capabilities adoption is optional to the user, as their application could run on \sys unchanged.
And even without offloading some system's part of the model to \sys, their performance would still be improved because of the \emph{ML for OS} part of the MLaaS subsystem.
Regarding the GPU virtualization software, since it is integrated as a kernel module, it requires no kernel recompilation or reboot.
Even updates in its code will be applied as simply as unloading and reloading the corresponding kernel module.



\section{Conclusion}
\label{sec:conclusion}
This paper introduced \sys, an ML-specialized OS that rethinks the kernel architecture to improve ML application performance in virtualized environments.
Driven by the GPU passthrough performance of recent workloads on newer GPU hardware, \sys proposes an approach to full GPU virtualization that follows the hardware-assisted CPU virtualization logic.

\bibliographystyle{ACM-Reference-Format}
\bibliography{paper}


\begin{thebibliography}{70}


\ifx \showCODEN    \undefined \def \showCODEN     #1{\unskip}     \fi
\ifx \showISBNx    \undefined \def \showISBNx     #1{\unskip}     \fi
\ifx \showISBNxiii \undefined \def \showISBNxiii  #1{\unskip}     \fi
\ifx \showISSN     \undefined \def \showISSN      #1{\unskip}     \fi
\ifx \showLCCN     \undefined \def \showLCCN      #1{\unskip}     \fi
\ifx \shownote     \undefined \def \shownote      #1{#1}          \fi
\ifx \showarticletitle \undefined \def \showarticletitle #1{#1}   \fi
\ifx \showURL      \undefined \def \showURL       {\relax}        \fi
\providecommand\bibfield[2]{#2}
\providecommand\bibinfo[2]{#2}
\providecommand\natexlab[1]{#1}
\providecommand\showeprint[2][]{arXiv:#2}

\bibitem[aws({[n.\,d.]})]%
        {aws-instance-pricing}
 \bibinfo{year}{[n.\,d.]}\natexlab{}.
\newblock \bibinfo{title}{Amazon EC2 P4 Instances}.
\newblock \bibinfo{howpublished}{\url{https://aws.amazon.com/fr/ec2/pricing/}}.
\newblock
\newblock
\shownote{Accessed: May 20, 2025}.


\bibitem[kit({[n.\,d.]})]%
        {kits19}
 \bibinfo{year}{[n.\,d.]}\natexlab{}.
\newblock \bibinfo{title}{KiTS19 Challenge Dataset}.
\newblock
  \bibinfo{howpublished}{\url{https://kits19.grand-challenge.org/data/}}.
\newblock
\newblock
\shownote{Accessed: [Jan 12, 2025]}.


\bibitem[num({[n.\,d.]})]%
        {numpy}
 \bibinfo{year}{[n.\,d.]}\natexlab{}.
\newblock \bibinfo{title}{NumPy - The fundamental package for scientific
  computing with Python}.
\newblock \bibinfo{howpublished}{\url{https://numpy.org/}}.
\newblock
\newblock
\shownote{Accessed: May 11, 2025}.


\bibitem[dal({[n.\,d.]})]%
        {dali}
 \bibinfo{year}{[n.\,d.]}\natexlab{}.
\newblock \bibinfo{title}{NVIDIA Data Loading Library (DALI)}.
\newblock \bibinfo{howpublished}{\url{https://developer.nvidia.com/dali}}.
\newblock
\newblock
\shownote{Accessed: May 5, 2025}.


\bibitem[pan({[n.\,d.]})]%
        {pandas}
 \bibinfo{year}{[n.\,d.]}\natexlab{}.
\newblock \bibinfo{title}{Pandas: powerful Python data analysis toolkit}.
\newblock \bibinfo{howpublished}{\url{https://pypi.org/project/pandas/}}.
\newblock
\newblock
\shownote{Accessed: May 11, 2025}.


\bibitem[sci({[n.\,d.]})]%
        {scikit}
 \bibinfo{year}{[n.\,d.]}\natexlab{}.
\newblock \bibinfo{title}{Scikit-learn - Machine Learning in Python}.
\newblock \bibinfo{howpublished}{\url{https://scikit-learn.org/stable/}}.
\newblock
\newblock
\shownote{Accessed: May 11, 2025}.


\bibitem[pri({[n.\,d.]})]%
        {princeton-os-cloud}
 \bibinfo{year}{[n.\,d.]}\natexlab{}.
\newblock \bibinfo{title}{The Top 3 Operating System in 2022; Linux, Windows,
  and Solaris}.
\newblock
  \bibinfo{howpublished}{\url{https://princetonits.com/blog/operating-system/the-top-3-operating-system-in-2022-linux-windows-and-solaris/}}.
\newblock
\newblock
\shownote{Accessed: May 19, 2025}.


\bibitem[wik({[n.\,d.]})]%
        {wiki-os-cloud}
 \bibinfo{year}{[n.\,d.]}\natexlab{}.
\newblock \bibinfo{title}{Usage share of operating systems}.
\newblock
  \bibinfo{howpublished}{\url{https://en.wikipedia.org/wiki/Usage_share_of_operating_systems}}.
\newblock
\newblock
\shownote{Accessed: May 19, 2025}.


\bibitem[nvm(2020)]%
        {nvml}
 \bibinfo{year}{2020}\natexlab{}.
\newblock \bibinfo{booktitle}{\emph{{NVIDIA Managemnet Library (NVML)}}}.
\newblock
\urldef\tempurl%
\url{https://developer.nvidia.com/management-library-nvml}
\showURL{%
\tempurl}
\newblock
\shownote{Accessed: May 11, 2025}.


\bibitem[Ahmed et~al\mbox{.}(2019)]%
        {ml.net}
\bibfield{author}{\bibinfo{person}{Zeeshan Ahmed}, \bibinfo{person}{Saeed
  Amizadeh}, \bibinfo{person}{Mikhail Bilenko}, \bibinfo{person}{Rogan Carr},
  \bibinfo{person}{Wei-Sheng Chin}, \bibinfo{person}{Yael Dekel},
  \bibinfo{person}{Xavier Dupre}, \bibinfo{person}{Vadim Eksarevskiy},
  \bibinfo{person}{Senja Filipi}, \bibinfo{person}{Tom Finley},
  \bibinfo{person}{Abhishek Goswami}, \bibinfo{person}{Monte Hoover},
  \bibinfo{person}{Scott Inglis}, \bibinfo{person}{Matteo Interlandi},
  \bibinfo{person}{Najeeb Kazmi}, \bibinfo{person}{Gleb Krivosheev},
  \bibinfo{person}{Pete Luferenko}, \bibinfo{person}{Ivan Matantsev},
  \bibinfo{person}{Sergiy Matusevych}, \bibinfo{person}{Shahab Moradi},
  \bibinfo{person}{Gani Nazirov}, \bibinfo{person}{Justin Ormont},
  \bibinfo{person}{Gal Oshri}, \bibinfo{person}{Artidoro Pagnoni},
  \bibinfo{person}{Jignesh Parmar}, \bibinfo{person}{Prabhat Roy},
  \bibinfo{person}{Mohammad~Zeeshan Siddiqui}, \bibinfo{person}{Markus Weimer},
  \bibinfo{person}{Shauheen Zahirazami}, {and} \bibinfo{person}{Yiwen Zhu}.}
  \bibinfo{year}{2019}\natexlab{}.
\newblock \showarticletitle{Machine Learning at Microsoft with ML.NET}. In
  \bibinfo{booktitle}{\emph{Proceedings of the ACM SIGKDD International
  Conference on Knowledge Discovery and Data Mining}}.
\newblock
\href{https://doi.org/10.1145/3292500.3330667}{doi:\nolinkurl{10.1145/3292500.3330667}}


\bibitem[AI({[n.\,d.]})]%
        {gemini}
\bibfield{author}{\bibinfo{person}{Google AI}.}
  \bibinfo{year}{[n.\,d.]}\natexlab{}.
\newblock \bibinfo{title}{Chat with Gemini to supercharge your creativity and
  productivity.}
\newblock
  \bibinfo{howpublished}{\url{https://store.google.com/intl/en/ideas/categories/ai/}}.
\newblock
\newblock
\shownote{Accessed: May 19, 2025}.


\bibitem[Android({[n.\,d.]})]%
        {android-ai}
\bibfield{author}{\bibinfo{person}{Android}.}
  \bibinfo{year}{[n.\,d.]}\natexlab{}.
\newblock \bibinfo{title}{Get the best of Google AI on Android.}
\newblock \bibinfo{howpublished}{\url{https://www.android.com/intl/en_ca/ai/}}.
\newblock
\newblock
\shownote{Accessed: May 19, 2025}.


\bibitem[Apple({[n.\,d.]})]%
        {apple-ai}
\bibfield{author}{\bibinfo{person}{Apple}.}
  \bibinfo{year}{[n.\,d.]}\natexlab{}.
\newblock \bibinfo{title}{Apple Intelligence - AI for the rest of us.}
\newblock
  \bibinfo{howpublished}{\url{https://www.apple.com/ca/apple-intelligence/}}.
\newblock
\newblock
\shownote{Accessed: May 19, 2025}.


\bibitem[Bugnion et~al\mbox{.}(2013)]%
        {Bugnion:2013:LSS:2490483.2490506}
\bibfield{author}{\bibinfo{person}{Edouard Bugnion}, \bibinfo{person}{Vitaly
  Chipounov}, {and} \bibinfo{person}{George Candea}.}
  \bibinfo{year}{2013}\natexlab{}.
\newblock \showarticletitle{Lightweight Snapshots and System-level
  Backtracking}. In \bibinfo{booktitle}{\emph{Proceedings of the 14th USENIX
  Conference on Hot Topics in Operating Systems}} (Santa Ana Pueblo, New
  Mexcio) \emph{(\bibinfo{series}{HotOS'13})}. \bibinfo{publisher}{USENIX
  Association}, \bibinfo{address}{Berkeley, CA, USA}, \bibinfo{pages}{23--23}.
\newblock
\urldef\tempurl%
\url{http://dl.acm.org/citation.cfm?id=2490483.2490506}
\showURL{%
\tempurl}


\bibitem[\c{C}i\c{c}ek et~al\mbox{.}(2016)]%
        {10.1007/978-3-319-46723-8_49}
\bibfield{author}{\bibinfo{person}{\"{O}zg\"{u}n \c{C}i\c{c}ek},
  \bibinfo{person}{Ahmed Abdulkadir}, \bibinfo{person}{Soeren~S. Lienkamp},
  \bibinfo{person}{Thomas Brox}, {and} \bibinfo{person}{Olaf Ronneberger}.}
  \bibinfo{year}{2016}\natexlab{}.
\newblock \showarticletitle{3D U-Net: Learning Dense Volumetric Segmentation
  from Sparse Annotation}.
\newblock


\bibitem[Chen et~al\mbox{.}(2020)]%
        {10.1145/3409963.3410492}
\bibfield{author}{\bibinfo{person}{Jingde Chen}, \bibinfo{person}{Subho~S.
  Banerjee}, \bibinfo{person}{Zbigniew~T. Kalbarczyk}, {and}
  \bibinfo{person}{Ravishankar~K. Iyer}.} \bibinfo{year}{2020}\natexlab{}.
\newblock \showarticletitle{Machine learning for load balancing in the Linux
  kernel}. In \bibinfo{booktitle}{\emph{Proceedings of the 11th ACM SIGOPS
  Asia-Pacific Workshop on Systems}}.
\newblock
\href{https://doi.org/10.1145/3409963.3410492}{doi:\nolinkurl{10.1145/3409963.3410492}}


\bibitem[Daube-Witherspoon et~al\mbox{.}(2001)]%
        {910827}
\bibfield{author}{\bibinfo{person}{M.E. Daube-Witherspoon}, \bibinfo{person}{S.
  Matej}, \bibinfo{person}{J.S. Karp}, {and} \bibinfo{person}{R.M. Lewitt}.}
  \bibinfo{year}{2001}\natexlab{}.
\newblock \showarticletitle{Application of the row action maximum likelihood
  algorithm with spherical basis functions to clinical PET imaging}.
\newblock \bibinfo{journal}{\emph{IEEE Transactions on Nuclear Science}}
  \bibinfo{volume}{48}, \bibinfo{number}{1} (\bibinfo{year}{2001}),
  \bibinfo{pages}{24--30}.
\newblock
\href{https://doi.org/10.1109/23.910827}{doi:\nolinkurl{10.1109/23.910827}}


\bibitem[Docker({[n.\,d.]})]%
        {nvidia-docker}
\bibfield{author}{\bibinfo{person}{Docker}.}
  \bibinfo{year}{[n.\,d.]}\natexlab{}.
\newblock \bibinfo{title}{NVIDIA Docker: GPU Server Application Deployment Made
  Easy}.
\newblock
  \bibinfo{howpublished}{\url{https://developer.nvidia.com/blog/nvidia-docker-gpu-server-application-deployment-made-easy/}}.
\newblock
\newblock
\shownote{Accessed: May 24, 2025}.


\bibitem[Doudali et~al\mbox{.}(2019)]%
        {kleio}
\bibfield{author}{\bibinfo{person}{Thaleia~Dimitra Doudali},
  \bibinfo{person}{Sergey Blagodurov}, \bibinfo{person}{Abhinav Vishnu},
  \bibinfo{person}{Sudhanva Gurumurthi}, {and} \bibinfo{person}{Ada
  Gavrilovska}.} \bibinfo{year}{2019}\natexlab{}.
\newblock \showarticletitle{Kleio: A Hybrid Memory Page Scheduler with Machine
  Intelligence}. In \bibinfo{booktitle}{\emph{Proceedings of the 28th
  International Symposium on High-Performance Parallel and Distributed
  Computing}}.
\newblock
\href{https://doi.org/10.1145/3307681.3325398}{doi:\nolinkurl{10.1145/3307681.3325398}}


\bibitem[Dowty and Sugerman(2009)]%
        {10.1145/1618525.1618534}
\bibfield{author}{\bibinfo{person}{Micah Dowty} {and} \bibinfo{person}{Jeremy
  Sugerman}.} \bibinfo{year}{2009}\natexlab{}.
\newblock \showarticletitle{GPU virtualization on VMware's hosted I/O
  architecture}.
\newblock \bibinfo{journal}{\emph{SIGOPS Opererating Systems Review}}
  (\bibinfo{year}{2009}).
\newblock
\href{https://doi.org/10.1145/1618525.1618534}{doi:\nolinkurl{10.1145/1618525.1618534}}


\bibitem[Fedorova et~al\mbox{.}(2007)]%
        {Fedorova2007OperatingSS}
\bibfield{author}{\bibinfo{person}{Alexandra Fedorova}, \bibinfo{person}{David
  Vengerov}, \bibinfo{person}{David Vengerov}, {and} \bibinfo{person}{Daniel
  Doucette}.} \bibinfo{year}{2007}\natexlab{}.
\newblock \showarticletitle{Operating System Scheduling On Heterogeneous Core
  Systems}.
\newblock
\urldef\tempurl%
\url{https://api.semanticscholar.org/CorpusID:14823905}
\showURL{%
\tempurl}


\bibitem[Fingler et~al\mbox{.}(2023)]%
        {lake}
\bibfield{author}{\bibinfo{person}{Henrique Fingler}, \bibinfo{person}{Isha
  Tarte}, \bibinfo{person}{Hangchen Yu}, \bibinfo{person}{Ariel Szekely},
  \bibinfo{person}{Bodun Hu}, \bibinfo{person}{Aditya Akella}, {and}
  \bibinfo{person}{Christopher~J. Rossbach}.} \bibinfo{year}{2023}\natexlab{}.
\newblock \showarticletitle{Towards a Machine Learning-Assisted Kernel with
  LAKE}. In \bibinfo{booktitle}{\emph{Proceedings of the 28th ACM International
  Conference on Architectural Support for Programming Languages and Operating
  Systems, Volume 2}}.
\newblock
\href{https://doi.org/10.1145/3575693.3575697}{doi:\nolinkurl{10.1145/3575693.3575697}}


\bibitem[G and Roogi({[n.\,d.]})]%
        {9488982}
\bibfield{author}{\bibinfo{person}{Chitralekha G} {and}
  \bibinfo{person}{Jyoti~M Roogi}.} \bibinfo{year}{[n.\,d.]}\natexlab{}.
\newblock \showarticletitle{A Quick Review of ML Algorithms}. In
  \bibinfo{booktitle}{\emph{2021 6th International Conference on Communication
  and Electronics Systems (ICCES)}}.
\newblock
\href{https://doi.org/10.1109/ICCES51350.2021.9488982}{doi:\nolinkurl{10.1109/ICCES51350.2021.9488982}}


\bibitem[Ganfure et~al\mbox{.}(2020)]%
        {deepprefetch}
\bibfield{author}{\bibinfo{person}{Gaddisa~Olani Ganfure},
  \bibinfo{person}{Chun-Feng Wu}, \bibinfo{person}{Yuan-Hao Chang}, {and}
  \bibinfo{person}{Wei-Kuan Shih}.} \bibinfo{year}{2020}\natexlab{}.
\newblock \showarticletitle{DeepPrefetcher: A Deep Learning Framework for Data
  Prefetching in Flash Storage Devices}.
\newblock \bibinfo{journal}{\emph{IEEE Transactions on Computer-Aided Design of
  Integrated Circuits and Systems}} (\bibinfo{year}{2020}).
\newblock
\href{https://doi.org/10.1109/TCAD.2020.3012173}{doi:\nolinkurl{10.1109/TCAD.2020.3012173}}


\bibitem[Goodarzy et~al\mbox{.}(2021)]%
        {smartos}
\bibfield{author}{\bibinfo{person}{Sepideh Goodarzy}, \bibinfo{person}{Maziyar
  Nazari}, \bibinfo{person}{Richard Han}, \bibinfo{person}{Eric Keller}, {and}
  \bibinfo{person}{Eric Rozner}.} \bibinfo{year}{2021}\natexlab{}.
\newblock \showarticletitle{SmartOS: towards automated learning and
  user-adaptive resource allocation in operating systems}. In
  \bibinfo{booktitle}{\emph{Proceedings of the 12th ACM SIGOPS Asia-Pacific
  Workshop on Systems}}.
\newblock
\href{https://doi.org/10.1145/3476886.3477519}{doi:\nolinkurl{10.1145/3476886.3477519}}


\bibitem[Gorgolewski et~al\mbox{.}(2011)]%
        {10.3389/fninf.2011.00013}
\bibfield{author}{\bibinfo{person}{Krzysztof Gorgolewski},
  \bibinfo{person}{Christopher Burns}, \bibinfo{person}{Cindee Madison},
  \bibinfo{person}{Dav Clark}, \bibinfo{person}{Yaroslav Halchenko},
  \bibinfo{person}{Michael Waskom}, {and} \bibinfo{person}{Satrajit Ghosh}.}
  \bibinfo{year}{2011}\natexlab{}.
\newblock \showarticletitle{Nipype: A Flexible, Lightweight and Extensible
  Neuroimaging Data Processing Framework in Python}.
\newblock \bibinfo{journal}{\emph{Frontiers in Neuroinformatics}}
  \bibinfo{volume}{5} (\bibinfo{year}{2011}).
\newblock
\showISSN{1662-5196}
\href{https://doi.org/10.3389/fninf.2011.00013}{doi:\nolinkurl{10.3389/fninf.2011.00013}}


\bibitem[Graur et~al\mbox{.}(2022)]%
        {cachew}
\bibfield{author}{\bibinfo{person}{Dan Graur}, \bibinfo{person}{Damien Aymon},
  \bibinfo{person}{Dan Kluser}, \bibinfo{person}{Tanguy Albrici},
  \bibinfo{person}{Chandramohan~A Thekkath}, {and} \bibinfo{person}{Ana
  Klimovic}.} \bibinfo{year}{2022}\natexlab{}.
\newblock \showarticletitle{Cachew: Machine Learning Input Data Processing As A
  Service}. In \bibinfo{booktitle}{\emph{Proceedings of USENIX ATC 22}}.
\newblock


\bibitem[Graur et~al\mbox{.}(2024)]%
        {pecan}
\bibfield{author}{\bibinfo{person}{Dan Graur}, \bibinfo{person}{Oto Mraz},
  \bibinfo{person}{Muyu Li}, \bibinfo{person}{Sepehr Pourghannad},
  \bibinfo{person}{Chandramohan~A. Thekkath}, {and} \bibinfo{person}{Ana
  Klimovic}.} \bibinfo{year}{2024}\natexlab{}.
\newblock \showarticletitle{Pecan: {Cost-Efficient} {ML} Data Preprocessing
  with Automatic Transformation Ordering and Hybrid Placement}. In
  \bibinfo{booktitle}{\emph{USENIX ATC}}.
\newblock


\bibitem[Habibi et~al\mbox{.}(2024)]%
        {habibi2024aimldriven}
\bibfield{author}{\bibinfo{person}{Mohammad~Asif Habibi}, \bibinfo{person}{Bin
  Han}, \bibinfo{person}{Merve Saimler}, \bibinfo{person}{Ignacio~Labrador
  Pavon}, {and} \bibinfo{person}{Hans~D. Schotten}.}
  \bibinfo{year}{2024}\natexlab{}.
\newblock \bibinfo{title}{Towards an AI/ML-driven SMO Framework in O-RAN:
  Scenarios, Solutions, and Challenges}.
\newblock
\urldef\tempurl%
\url{https://arxiv.org/abs/2409.05092}
\showURL{%
\tempurl}


\bibitem[Hao et~al\mbox{.}(2020)]%
        {linnos}
\bibfield{author}{\bibinfo{person}{Mingzhe Hao}, \bibinfo{person}{Levent
  Toksoz}, \bibinfo{person}{Nanqinqin Li}, \bibinfo{person}{Edward~Edberg
  Halim}, \bibinfo{person}{Henry Hoffmann}, {and} \bibinfo{person}{Haryadi~S.
  Gunawi}.} \bibinfo{year}{2020}\natexlab{}.
\newblock \showarticletitle{{LinnOS}: Predictability on Unpredictable Flash
  Storage with a Light Neural Network}. In \bibinfo{booktitle}{\emph{14th
  USENIX Symposium on Operating Systems Design and Implementation}}.
\newblock


\bibitem[Hawkins et~al\mbox{.}(2010)]%
        {10.1021/ci100031x}
\bibfield{author}{\bibinfo{person}{Paul Hawkins}, \bibinfo{person}{Geoff
  Skillman}, \bibinfo{person}{Gregory Warren}, \bibinfo{person}{Benjamin
  Ellingson}, {and} \bibinfo{person}{Matthew Stahl}.}
  \bibinfo{year}{2010}\natexlab{}.
\newblock \showarticletitle{Conformer Generation with OMEGA: Algorithm and
  Validation Using High Quality Structures from the Protein Databank and
  Cambridge Structural Database}.
\newblock \bibinfo{journal}{\emph{Journal of chemical information and
  modeling}}  \bibinfo{volume}{50} (\bibinfo{date}{03} \bibinfo{year}{2010}),
  \bibinfo{pages}{572--84}.
\newblock
\href{https://doi.org/10.1021/ci100031x}{doi:\nolinkurl{10.1021/ci100031x}}


\bibitem[Heller et~al\mbox{.}(2020)]%
        {heller2019kits19}
\bibfield{author}{\bibinfo{person}{Nicholas Heller}, \bibinfo{person}{Niranjan
  Sathianathen}, \bibinfo{person}{Arveen Kalapara}, \bibinfo{person}{Edward
  Walczak}, \bibinfo{person}{Keenan Moore}, \bibinfo{person}{Heather
  Kaluzniak}, \bibinfo{person}{Joel Rosenberg}, \bibinfo{person}{Paul Blake},
  \bibinfo{person}{Zachary Rengel}, \bibinfo{person}{Makinna Oestreich},
  \bibinfo{person}{Joshua Dean}, \bibinfo{person}{Michael Tradewell},
  \bibinfo{person}{Aneri Shah}, \bibinfo{person}{Resha Tejpaul},
  \bibinfo{person}{Zachary Edgerton}, \bibinfo{person}{Matthew Peterson},
  \bibinfo{person}{Shaneabbas Raza}, \bibinfo{person}{Subodh Regmi},
  \bibinfo{person}{Nikolaos Papanikolopoulos}, {and}
  \bibinfo{person}{Christopher Weight}.} \bibinfo{year}{2020}\natexlab{}.
\newblock \bibinfo{title}{The KiTS19 Challenge Data: 300 Kidney Tumor Cases
  with Clinical Context, CT Semantic Segmentations, and Surgical Outcomes}.
\newblock
\showeprint{1904.00445}
\urldef\tempurl%
\url{https://arxiv.org/abs/1904.00445}
\showURL{%
\tempurl}


\bibitem[Hong et~al\mbox{.}(2017)]%
        {10.1145/3068281}
\bibfield{author}{\bibinfo{person}{Cheol-Ho Hong}, \bibinfo{person}{Ivor
  Spence}, {and} \bibinfo{person}{Dimitrios~S. Nikolopoulos}.}
  \bibinfo{year}{2017}\natexlab{}.
\newblock \showarticletitle{GPU Virtualization and Scheduling Methods: A
  Comprehensive Survey}.
\newblock \bibinfo{journal}{\emph{Comput. Surveys}} (\bibinfo{year}{2017}).
\newblock
\href{https://doi.org/10.1145/3068281}{doi:\nolinkurl{10.1145/3068281}}


\bibitem[Hou et~al\mbox{.}(2016)]%
        {deepmaldroid}
\bibfield{author}{\bibinfo{person}{Shifu Hou}, \bibinfo{person}{Aaron Saas},
  \bibinfo{person}{Lifei Chen}, {and} \bibinfo{person}{Yanfang Ye}.}
  \bibinfo{year}{2016}\natexlab{}.
\newblock \showarticletitle{Deep4MalDroid: A Deep Learning Framework for
  Android Malware Detection Based on Linux Kernel System Call Graphs}. In
  \bibinfo{booktitle}{\emph{2016 IEEE/WIC/ACM International Conference on Web
  Intelligence Workshops}}.
\newblock
\href{https://doi.org/10.1109/WIW.2016.040}{doi:\nolinkurl{10.1109/WIW.2016.040}}


\bibitem[Ibrahim and Oliker(2022)]%
        {9820645}
\bibfield{author}{\bibinfo{person}{Khaled~Z. Ibrahim} {and}
  \bibinfo{person}{Leonid Oliker}.} \bibinfo{year}{2022}\natexlab{}.
\newblock \showarticletitle{Preprocessing Pipeline Optimization for Scientific
  Deep Learning Workloads}. In \bibinfo{booktitle}{\emph{IEEE International
  Parallel and Distributed Processing Symposium (IPDPS)}}.
\newblock
\href{https://doi.org/10.1109/IPDPS53621.2022.00112}{doi:\nolinkurl{10.1109/IPDPS53621.2022.00112}}


\bibitem[Kang et~al\mbox{.}(2017)]%
        {convgpu}
\bibfield{author}{\bibinfo{person}{Daeyoun Kang}, \bibinfo{person}{Tae~Joon
  Jun}, \bibinfo{person}{Dohyeun Kim}, \bibinfo{person}{Jaewook Kim}, {and}
  \bibinfo{person}{Daeyoung Kim}.} \bibinfo{year}{2017}\natexlab{}.
\newblock \showarticletitle{ConVGPU: GPU Management Middleware in Container
  Based Virtualized Environment}. In \bibinfo{booktitle}{\emph{2017 IEEE
  International Conference on Cluster Computing (CLUSTER)}}.
\newblock
\href{https://doi.org/10.1109/CLUSTER.2017.17}{doi:\nolinkurl{10.1109/CLUSTER.2017.17}}


\bibitem[Karypis and Kumar(1998)]%
        {doi:10.1137/S1064827595287997}
\bibfield{author}{\bibinfo{person}{George Karypis} {and} \bibinfo{person}{Vipin
  Kumar}.} \bibinfo{year}{1998}\natexlab{}.
\newblock \showarticletitle{A Fast and High Quality Multilevel Scheme for
  Partitioning Irregular Graphs}.
\newblock \bibinfo{journal}{\emph{SIAM Journal on Scientific Computing}}
  \bibinfo{volume}{20}, \bibinfo{number}{1} (\bibinfo{year}{1998}),
  \bibinfo{pages}{359--392}.
\newblock
\href{https://doi.org/10.1137/S1064827595287997}{doi:\nolinkurl{10.1137/S1064827595287997}}
\showeprint{https://doi.org/10.1137/S1064827595287997}


\bibitem[Kotsiantis et~al\mbox{.}(2006a)]%
        {processing-sl}
\bibfield{author}{\bibinfo{person}{Sotiris Kotsiantis},
  \bibinfo{person}{Dimitris Kanellopoulos}, {and} \bibinfo{person}{P.
  Pintelas}.} \bibinfo{year}{2006}\natexlab{a}.
\newblock \showarticletitle{Data Preprocessing for Supervised Learning}.
\newblock \bibinfo{journal}{\emph{International Journal of Computer Science}}
  \bibinfo{volume}{1} (\bibinfo{date}{01} \bibinfo{year}{2006}),
  \bibinfo{pages}{111--117}.
\newblock


\bibitem[Kotsiantis et~al\mbox{.}(2006b)]%
        {10.5281/zenodo.1082415}
\bibfield{author}{\bibinfo{person}{Sotiris Kotsiantis},
  \bibinfo{person}{Dimitris Kanellopoulos}, {and} \bibinfo{person}{P.
  Pintelas}.} \bibinfo{year}{2006}\natexlab{b}.
\newblock \showarticletitle{Data Preprocessing for Supervised Learning}.
\newblock \bibinfo{journal}{\emph{International Journal of Computer Science}}
  (\bibinfo{year}{2006}).
\newblock


\bibitem[Laga et~al\mbox{.}(2016)]%
        {lynx}
\bibfield{author}{\bibinfo{person}{Arezki Laga}, \bibinfo{person}{Jalil
  Boukhobza}, \bibinfo{person}{Michel Koskas}, {and} \bibinfo{person}{Frank
  Singhoff}.} \bibinfo{year}{2016}\natexlab{}.
\newblock \showarticletitle{Lynx: a learning linux prefetching mechanism for
  SSD performance model}. In \bibinfo{booktitle}{\emph{2016 5th Non-Volatile
  Memory Systems and Applications Symposium}}.
\newblock
\href{https://doi.org/10.1109/NVMSA.2016.7547186}{doi:\nolinkurl{10.1109/NVMSA.2016.7547186}}


\bibitem[Lee et~al\mbox{.}(2016)]%
        {vadi}
\bibfield{author}{\bibinfo{person}{Chiyoung Lee}, \bibinfo{person}{Se-Won Kim},
  {and} \bibinfo{person}{Chuck Yoo}.} \bibinfo{year}{2016}\natexlab{}.
\newblock \showarticletitle{VADI: GPU Virtualization for an Automotive
  Platform}.
\newblock \bibinfo{journal}{\emph{IEEE Transactions on Industrial Informatics}}
  (\bibinfo{year}{2016}).
\newblock


\bibitem[Lin et~al\mbox{.}(2014)]%
        {cocodataset}
\bibfield{author}{\bibinfo{person}{Tsung{-}Yi Lin}, \bibinfo{person}{Michael
  Maire}, \bibinfo{person}{Serge~J. Belongie}, \bibinfo{person}{Lubomir~D.
  Bourdev}, \bibinfo{person}{Ross~B. Girshick}, \bibinfo{person}{James Hays},
  \bibinfo{person}{Pietro Perona}, \bibinfo{person}{Deva Ramanan},
  \bibinfo{person}{Piotr Doll{'{a} }r}, {and} \bibinfo{person}{C.~Lawrence
  Zitnick}.} \bibinfo{year}{2014}\natexlab{}.
\newblock \showarticletitle{Microsoft {COCO:} Common Objects in Context}.
\newblock \bibinfo{journal}{\emph{CoRR}} (\bibinfo{year}{2014}).
\newblock
\urldef\tempurl%
\url{http://arxiv.org/abs/1405.0312}
\showURL{%
\tempurl}


\bibitem[Maas et~al\mbox{.}(2020)]%
        {10.1145/3373376.3378525}
\bibfield{author}{\bibinfo{person}{Martin Maas}, \bibinfo{person}{David~G.
  Andersen}, \bibinfo{person}{Michael Isard}, \bibinfo{person}{Mohammad~Mahdi
  Javanmard}, \bibinfo{person}{Kathryn~S. McKinley}, {and}
  \bibinfo{person}{Colin Raffel}.} \bibinfo{year}{2020}\natexlab{}.
\newblock \showarticletitle{Learning-based Memory Allocation for C++ Server
  Workloads}. In \bibinfo{booktitle}{\emph{Proceedings of the Twenty-Fifth
  International Conference on Architectural Support for Programming Languages
  and Operating Systems}}.
\newblock
\href{https://doi.org/10.1145/3373376.3378525}{doi:\nolinkurl{10.1145/3373376.3378525}}


\bibitem[Maetschke et~al\mbox{.}(2017)]%
        {nuts-flow/ml}
\bibfield{author}{\bibinfo{person}{Stefan Maetschke},
  \bibinfo{person}{Ruwan~Bandara Tennakoon}, \bibinfo{person}{Christian
  Vecchiola}, {and} \bibinfo{person}{Rahil Garnavi}.}
  \bibinfo{year}{2017}\natexlab{}.
\newblock \showarticletitle{nuts-flow/ml: data pre-processing for deep
  learning}.
\newblock  (\bibinfo{year}{2017}).
\newblock
\showeprint[arXiv]{1708.06046}
\urldef\tempurl%
\url{http://arxiv.org/abs/1708.06046}
\showURL{%
\tempurl}


\bibitem[Makino et~al\mbox{.}(2019)]%
        {rnnt}
\bibfield{author}{\bibinfo{person}{Takaki Makino}, \bibinfo{person}{Hank Liao},
  \bibinfo{person}{Yannis Assael}, \bibinfo{person}{Brendan Shillingford},
  \bibinfo{person}{Basilio Garcia}, \bibinfo{person}{Otavio Braga}, {and}
  \bibinfo{person}{Olivier Siohan}.} \bibinfo{year}{2019}\natexlab{}.
\newblock \showarticletitle{Recurrent Neural Network Transducer for
  Audio-Visual Speech Recognition}. In \bibinfo{booktitle}{\emph{IEEE Automatic
  Speech Recognition and Understanding Workshop}}.
\newblock


\bibitem[Massa and Girshick({[n.\,d.]})]%
        {massa2018mrcnn}
\bibfield{author}{\bibinfo{person}{Francisco Massa} {and} \bibinfo{person}{Ross
  Girshick}.} \bibinfo{year}{[n.\,d.]}\natexlab{}.
\newblock \bibinfo{title}{{maskrnn-benchmark: Fast, modular reference
  implementation of Instance Segmentation and Object Detection algorithms in
  PyTorch}}.
\newblock
  \bibinfo{howpublished}{\url{https://github.com/facebookresearch/maskrcnn-benchmark}}.
\newblock
\newblock
\shownote{Accessed: May 11, 2025}.


\bibitem[Mattson et~al\mbox{.}(2020)]%
        {mattson2020mlperf}
\bibfield{author}{\bibinfo{person}{Peter Mattson}, \bibinfo{person}{Christine
  Cheng}, \bibinfo{person}{Gregory Diamos}, \bibinfo{person}{Cody Coleman},
  \bibinfo{person}{Paulius Micikevicius}, \bibinfo{person}{David Patterson},
  \bibinfo{person}{Hanlin Tang}, \bibinfo{person}{Gu-Yeon Wei},
  \bibinfo{person}{Peter Bailis}, \bibinfo{person}{Victor Bittorf},
  {et~al\mbox{.}}} \bibinfo{year}{2020}\natexlab{}.
\newblock \showarticletitle{{MlPerf Training Benchmark}}.
\newblock \bibinfo{journal}{\emph{Proceedings of Machine Learning and Systems}}
  (\bibinfo{year}{2020}).
\newblock


\bibitem[{MLCommons}({[n.\,d.]})]%
        {mlcommons-github}
\bibfield{author}{\bibinfo{person}{{MLCommons}}.}
  \bibinfo{year}{[n.\,d.]}\natexlab{}.
\newblock \bibinfo{title}{{MLPerf Benchmarking Suite - PyTorch implementation
  for image segmentation}}.
\newblock
  \bibinfo{howpublished}{\url{https://github.com/mlcommons/training/tree/master/image_segmentation/pytorch}}.
\newblock
\newblock
\shownote{Accessed: [May 5, 2025]}.


\bibitem[Murray et~al\mbox{.}(2021)]%
        {tfdata}
\bibfield{author}{\bibinfo{person}{Derek~Gordon Murray},
  \bibinfo{person}{Jir{\'i} Simsa}, \bibinfo{person}{Ana Klimovic}, {and}
  \bibinfo{person}{Ihor Indyk}.} \bibinfo{year}{2021}\natexlab{}.
\newblock \showarticletitle{tf.data: A Machine Learning Data Processing
  Framework}.
\newblock \bibinfo{journal}{\emph{Proceedings of the VLDB Endowment}}
  (\bibinfo{year}{2021}).
\newblock


\bibitem[Musse and Alamro(2016)]%
        {7976640}
\bibfield{author}{\bibinfo{person}{Hodan~M. Musse} {and}
  \bibinfo{person}{Lama~A. Alamro}.} \bibinfo{year}{2016}\natexlab{}.
\newblock \showarticletitle{Cloud Computing: Architecture and Operating
  System}. In \bibinfo{booktitle}{\emph{Global Summit on Computer and
  Information Technology (GSCIT)}}.
\newblock
\href{https://doi.org/10.1109/GSCIT.2016.7}{doi:\nolinkurl{10.1109/GSCIT.2016.7}}


\bibitem[Negi and Kumar(2005)]%
        {4085157}
\bibfield{author}{\bibinfo{person}{Atul Negi} {and} \bibinfo{person}{P.~Kishore
  Kumar}.} \bibinfo{year}{2005}\natexlab{}.
\newblock \showarticletitle{Applying Machine Learning Techniques to Improve
  Linux Process Scheduling}. In \bibinfo{booktitle}{\emph{TENCON IEEE Region 10
  Conference}}.
\newblock
\href{https://doi.org/10.1109/TENCON.2005.300837}{doi:\nolinkurl{10.1109/TENCON.2005.300837}}


\bibitem[Nouaji et~al\mbox{.}(2024)]%
        {speedyloader}
\bibfield{author}{\bibinfo{person}{Rahma Nouaji}, \bibinfo{person}{Stella
  Bitchebe}, {and} \bibinfo{person}{Oana Balmau}.}
  \bibinfo{year}{2024}\natexlab{}.
\newblock \showarticletitle{SpeedyLoader: Efficient Pipelining of Data
  Preprocessing and Machine Learning Training}. In
  \bibinfo{booktitle}{\emph{Proceedings of the 4th Workshop on Machine Learning
  and Systems}}.
\newblock
\href{https://doi.org/10.1145/3642970.3655824}{doi:\nolinkurl{10.1145/3642970.3655824}}


\bibitem[NVIDIA({[n.\,d.]})]%
        {nvidiavgpu}
\bibfield{author}{\bibinfo{person}{NVIDIA}.}
  \bibinfo{year}{[n.\,d.]}\natexlab{}.
\newblock \bibinfo{title}{NVIDIA Virtual GPU (vGPU) Software}.
\newblock
  \bibinfo{howpublished}{\url{https://docs.nvidia.com/vgpu/index.html}}.
\newblock
\newblock
\shownote{Accessed: May 20, 2025}.


\bibitem[Panayotov et~al\mbox{.}(2015)]%
        {librispeech}
\bibfield{author}{\bibinfo{person}{Vassil Panayotov}, \bibinfo{person}{Guoguo
  Chen}, \bibinfo{person}{Daniel Povey}, {and} \bibinfo{person}{Sanjeev
  Khudanpur}.} \bibinfo{year}{2015}\natexlab{}.
\newblock \showarticletitle{Librispeech: An ASR Corpus Based on Public Domain
  Audio Books}. In \bibinfo{booktitle}{\emph{2015 IEEE International Conference
  on Acoustics, Speech and Signal Processing (ICASSP)}}.
\newblock


\bibitem[Pianese et~al\mbox{.}(2010)]%
        {5486552}
\bibfield{author}{\bibinfo{person}{Fabio Pianese}, \bibinfo{person}{Peter
  Bosch}, \bibinfo{person}{Alessandro Duminuco}, \bibinfo{person}{Nico
  Janssens}, \bibinfo{person}{Thanos Stathopoulos}, {and}
  \bibinfo{person}{Moritz Steiner}.} \bibinfo{year}{2010}\natexlab{}.
\newblock \showarticletitle{Toward a Cloud Operating System}. In
  \bibinfo{booktitle}{\emph{2010 IEEE/IFIP Network Operations and Management
  Symposium Workshops}}.
\newblock
\href{https://doi.org/10.1109/NOMSW.2010.5486552}{doi:\nolinkurl{10.1109/NOMSW.2010.5486552}}


\bibitem[Qiu et~al\mbox{.}(2021)]%
        {10.1145/3458336.3465288}
\bibfield{author}{\bibinfo{person}{Yiming Qiu}, \bibinfo{person}{Hongyi Liu},
  \bibinfo{person}{Thomas Anderson}, \bibinfo{person}{Yingyan Lin}, {and}
  \bibinfo{person}{Ang Chen}.} \bibinfo{year}{2021}\natexlab{}.
\newblock \showarticletitle{Toward reconfigurable kernel datapaths with learned
  optimizations}. In \bibinfo{booktitle}{\emph{Proceedings of the Workshop on
  Hot Topics in Operating Systems}}.
\newblock
\href{https://doi.org/10.1145/3458336.3465288}{doi:\nolinkurl{10.1145/3458336.3465288}}


\bibitem[Reddy et~al\mbox{.}(2020)]%
        {9036908}
\bibfield{author}{\bibinfo{person}{G.~Thippa Reddy},
  \bibinfo{person}{M.~Praveen~Kumar Reddy}, \bibinfo{person}{Kuruva
  Lakshmanna}, \bibinfo{person}{Rajesh Kaluri},
  \bibinfo{person}{Dharmendra~Singh Rajput}, \bibinfo{person}{Gautam
  Srivastava}, {and} \bibinfo{person}{Thar Baker}.}
  \bibinfo{year}{2020}\natexlab{}.
\newblock \showarticletitle{Analysis of Dimensionality Reduction Techniques on
  Big Data}.
\newblock \bibinfo{journal}{\emph{IEEE Access}} (\bibinfo{year}{2020}).
\newblock
\href{https://doi.org/10.1109/ACCESS.2020.2980942}{doi:\nolinkurl{10.1109/ACCESS.2020.2980942}}


\bibitem[Shea and Liu(2013)]%
        {6820614}
\bibfield{author}{\bibinfo{person}{Ryan Shea} {and} \bibinfo{person}{Jiangchuan
  Liu}.} \bibinfo{year}{2013}\natexlab{}.
\newblock \showarticletitle{On GPU pass-through performance for cloud gaming:
  Experiments and analysis}. In \bibinfo{booktitle}{\emph{2013 12th Annual
  Workshop on Network and Systems Support for Games (NetGames)}}.
\newblock
\href{https://doi.org/10.1109/NetGames.2013.6820614}{doi:\nolinkurl{10.1109/NetGames.2013.6820614}}


\bibitem[Sudarshan et~al\mbox{.}(2006)]%
        {rf-preporcessing}
\bibfield{author}{\bibinfo{person}{Pallav Sudarshan},
  \bibinfo{person}{Neelesh~B. Mehta}, \bibinfo{person}{Andreas~F. Molisch},
  {and} \bibinfo{person}{Jin Zhang}.} \bibinfo{year}{2006}\natexlab{}.
\newblock \showarticletitle{Channel Statistics-Based RF Pre-Processing with
  Antenna Selection}.
\newblock \bibinfo{journal}{\emph{IEEE Transactions on Wireless
  Communications}} (\bibinfo{year}{2006}).
\newblock
\href{https://doi.org/10.1109/TWC.2006.256973}{doi:\nolinkurl{10.1109/TWC.2006.256973}}


\bibitem[Suranauwarat and Taniguchi(2001)]%
        {10.1145/506084.506090}
\bibfield{author}{\bibinfo{person}{Sukanya Suranauwarat} {and}
  \bibinfo{person}{Hideo Taniguchi}.} \bibinfo{year}{2001}\natexlab{}.
\newblock \showarticletitle{The design, implementation and initial evaluation
  of an advanced knowledge-based process scheduler}.
\newblock \bibinfo{journal}{\emph{SIGOPS Operating Systems Reviews}}
  (\bibinfo{year}{2001}).
\newblock
\href{https://doi.org/10.1145/506084.506090}{doi:\nolinkurl{10.1145/506084.506090}}


\bibitem[Suzuki et~al\mbox{.}(2016)]%
        {7349172}
\bibfield{author}{\bibinfo{person}{Yusuke Suzuki}, \bibinfo{person}{Shinpei
  Kato}, \bibinfo{person}{Hiroshi Yamada}, {and} \bibinfo{person}{Kenji Kono}.}
  \bibinfo{year}{2016}\natexlab{}.
\newblock \showarticletitle{GPUvm: GPU Virtualization at the Hypervisor}.
\newblock \bibinfo{journal}{\emph{IEEE Trans. Comput.}} (\bibinfo{year}{2016}).
\newblock
\href{https://doi.org/10.1109/TC.2015.2506582}{doi:\nolinkurl{10.1109/TC.2015.2506582}}


\bibitem[Teguia et~al\mbox{.}(2024)]%
        {vpim}
\bibfield{author}{\bibinfo{person}{Dufy Teguia}, \bibinfo{person}{Jiaxuan
  Chen}, \bibinfo{person}{Stella Bitchebe}, \bibinfo{person}{Oana Balmau},
  {and} \bibinfo{person}{Alain Tchana}.} \bibinfo{year}{2024}\natexlab{}.
\newblock \showarticletitle{vPIM: Processing-in-Memory Virtualization}. In
  \bibinfo{booktitle}{\emph{Proceedings of the 25th International Middleware
  Conference}}.
\newblock
\href{https://doi.org/10.1145/3652892.3700782}{doi:\nolinkurl{10.1145/3652892.3700782}}


\bibitem[Tesla({[n.\,d.]})]%
        {tesla-autopilot}
\bibfield{author}{\bibinfo{person}{Tesla}.}
  \bibinfo{year}{[n.\,d.]}\natexlab{}.
\newblock \bibinfo{title}{Autopilot and Full Self-Driving (Supervised)}.
\newblock
  \bibinfo{howpublished}{\url{https://www.tesla.com/support/autopilot}}.
\newblock
\newblock
\shownote{Accessed: May 19, 2025}.


\bibitem[Um et~al\mbox{.}(2023)]%
        {fastflow}
\bibfield{author}{\bibinfo{person}{Taegeon Um}, \bibinfo{person}{Byungsoo Oh},
  \bibinfo{person}{Byeongchan Seo}, \bibinfo{person}{Minhyeok Kweun},
  \bibinfo{person}{Goeun Kim}, {and} \bibinfo{person}{Woo-Yeon Lee}.}
  \bibinfo{year}{2023}\natexlab{}.
\newblock \showarticletitle{Fastflow: Accelerating Deep Learning Model Training
  With Smart Offloading of Input Data Pipeline}.
\newblock \bibinfo{journal}{\emph{Proceedings of the VLDB Endowment}}
  (\bibinfo{year}{2023}).
\newblock


\bibitem[Wentzlaff et~al\mbox{.}(2010)]%
        {10.1145/1807128.1807132}
\bibfield{author}{\bibinfo{person}{David Wentzlaff}, \bibinfo{person}{Charles
  Gruenwald}, \bibinfo{person}{Nathan Beckmann}, \bibinfo{person}{Kevin
  Modzelewski}, \bibinfo{person}{Adam Belay}, \bibinfo{person}{Lamia Youseff},
  \bibinfo{person}{Jason Miller}, {and} \bibinfo{person}{Anant Agarwal}.}
  \bibinfo{year}{2010}\natexlab{}.
\newblock \showarticletitle{An operating system for multicore and clouds:
  mechanisms and implementation}. In \bibinfo{booktitle}{\emph{Proceedings of
  the 1st ACM Symposium on Cloud Computing}}.
\newblock
\href{https://doi.org/10.1145/1807128.1807132}{doi:\nolinkurl{10.1145/1807128.1807132}}


\bibitem[Wikipedia({[n.\,d.]})]%
        {apple-ai-wiki}
\bibfield{author}{\bibinfo{person}{Wikipedia}.}
  \bibinfo{year}{[n.\,d.]}\natexlab{}.
\newblock \bibinfo{title}{Apple Intelligence}.
\newblock
  \bibinfo{howpublished}{\url{https://en.wikipedia.org/wiki/Apple_Intelligence}}.
\newblock
\newblock
\shownote{Accessed: May 19, 2025}.


\bibitem[Xu et~al\mbox{.}(2019)]%
        {8664598}
\bibfield{author}{\bibinfo{person}{Zhiyuan Xu}, \bibinfo{person}{Jian Tang},
  \bibinfo{person}{Chengxiang Yin}, \bibinfo{person}{Yanzhi Wang}, {and}
  \bibinfo{person}{Guoliang Xue}.} \bibinfo{year}{2019}\natexlab{}.
\newblock \showarticletitle{Experience-Driven Congestion Control: When
  Multi-Path TCP Meets Deep Reinforcement Learning}.
\newblock \bibinfo{journal}{\emph{IEEE Journal on Selected Areas in
  Communications}} (\bibinfo{year}{2019}).
\newblock
\href{https://doi.org/10.1109/JSAC.2019.2904358}{doi:\nolinkurl{10.1109/JSAC.2019.2904358}}


\bibitem[Xue et~al\mbox{.}(2023)]%
        {vnpu}
\bibfield{author}{\bibinfo{person}{Yuqi Xue}, \bibinfo{person}{Yiqi Liu}, {and}
  \bibinfo{person}{Jian Huang}.} \bibinfo{year}{2023}\natexlab{}.
\newblock \showarticletitle{System Virtualization for Neural Processing Units}.
  In \bibinfo{booktitle}{\emph{Proceedings of the 19th Workshop on Hot Topics
  in Operating Systems}}.
\newblock
\href{https://doi.org/10.1145/3593856.3595912}{doi:\nolinkurl{10.1145/3593856.3595912}}


\bibitem[Yang et~al\mbox{.}(2012)]%
        {6427531}
\bibfield{author}{\bibinfo{person}{Chao-Tung Yang}, \bibinfo{person}{Hsien-Yi
  Wang}, \bibinfo{person}{Wei-Shen Ou}, \bibinfo{person}{Yu-Tso Liu}, {and}
  \bibinfo{person}{Ching-Hsien Hsu}.} \bibinfo{year}{2012}\natexlab{}.
\newblock \showarticletitle{On implementation of GPU virtualization using PCI
  pass-through}. In \bibinfo{booktitle}{\emph{4th IEEE International Conference
  on Cloud Computing Technology and Science Proceedings}}.
\newblock
\href{https://doi.org/10.1109/CloudCom.2012.6427531}{doi:\nolinkurl{10.1109/CloudCom.2012.6427531}}


\bibitem[Younge et~al\mbox{.}({[n.\,d.]})]%
        {6969471}
\bibfield{author}{\bibinfo{person}{Andrew~J. Younge},
  \bibinfo{person}{John~Paul Walters}, \bibinfo{person}{Stephen Crago}, {and}
  \bibinfo{person}{Geoffrey~C. Fox}.} \bibinfo{year}{[n.\,d.]}\natexlab{}.
\newblock \showarticletitle{Evaluating GPU Passthrough in Xen for High
  Performance Cloud Computing}. In \bibinfo{booktitle}{\emph{IEEE International
  Parallel and Distributed Processing Symposium Workshops}}.
\newblock
\href{https://doi.org/10.1109/IPDPSW.2014.97}{doi:\nolinkurl{10.1109/IPDPSW.2014.97}}


\end{thebibliography}
\end{document}